\documentclass{appolb}
\usepackage{epsfig}
\begin{document}
% \eqsec  % uncomment this line to get equations numbered by (sec.num)
\title{Energy dependence of elliptic flow of $\phi$-meson in STAR at RHIC%
\thanks{Presented at Strangeness in Quark Matter Conference}%
% you can use '\\' to break lines
}
\author{md. nasim ( for star collaboration)
\address{Variable Energy Cyclotron Centre , Kolkata-700064, India}
}

\maketitle
\begin{abstract}
We present the measurement of elliptic flow ($v_{2}$) as a function of
transverse momentum ($p_{T}$) for the
$\phi$-meson in Au+Au collisions at $\sqrt{s_{NN}}$ = 7.7 , 11.5 and
39 GeV. At low $p_{T}$ $\phi$-meson  $v_{2}$ decreases with decrease
in beam energy.
 The number of constituent quark ($n_{\rm{cq}}$) scaled $v_{2}$ for
$\phi$-meson vs ($m_{T} - m_{0}$)/$n_{\rm{cq}}$ follows similar trend as other
hadrons for $\sqrt{s_{NN}}$= 39 GeV  whereas  the $n_{\rm{cq}}$ scaled 
$v_{2}$ for $\phi$-mesons follows a different trend compared to the
other hadrons  at $\sqrt{s_{NN}}$= 11.5
GeV.
\end{abstract}
  
\section{Introduction}
Tha elliptic flow ($v_{2}$) measured in heavy-ion collision are believed to
arise because of the pressure gradients developed in the overlap
region of two nuclei
collide at nonzero impact parameters. According to hydrodynamical
model $v_{2}$ is an early time phenomenon and  sensitive to the
equation of state of the system formed in the collision~\cite{hydro}. So $v_{2}$
can be used as a probe for early system although its magnitude may
change due to later stage hadronic interactions. The $\phi$-meson, which is the
bound state of s and $\bar{s}$ quark, has small interaction cross-section
with hadrons~\cite{smallx}. So for the $\phi$-meson $v_{2}$, effect of later stage
hadronic interaction is small. Therefore elliptic flow of $\phi$-mesons can be used as a clean
probe to measure early time collectivity of the system created in heavy-ion collision. Recent results
from Relativistic Heavy Ion Collider (RHIC) on $v_{2} $ as
function of transverse momentum ($p_{T}$)  shows
that at low $p_{T}$ elliptic flow of identified
hadrons follows mass ordering (lower $v_{2}$ for heavier hadrons than that of
lighter hadrons) whereas at intermediate $p_{T}$ all
mesons and all baryons form  two different groups.
When $v_{2}$ and $p_{T}$ are scaled by number of constituent
quarks of the hadrons, the measured $v_{2}$ values are consistent with
each other as parton coalescence or recombination models
predicted~\cite{ncq1,ncq1a}. This observation, is known as
 number of constituents quark scaling (NCQ scaling). This effect has been interpreted as
 collectivity being developed at the partonic stage of the evolution
 of the system in heavy-ion collision~\cite{ncq2}. $\phi$-meson has
 mass (1.02 GeV/$c^{2}$)
 comparable to mass of the lightest baryons (protons,$\Lambda$s) and at
 the same time it is meson, so study of NCQ scaling of $\phi$-meson
 $v_{2}$ would be more appropriate to understand the
 collectivity at partonic level and coalescence as mechanism of hadronization at intermediate $p_{T}$.
One of the main goal of RHIC Beam Energy Scan (BES) program is to see NCQ
scaling of $v_{2}$ for all hadrons as function of energy. In this program 
$\phi$-mesons $v_{2}$ plays an important role. Because violation of NCQ
scaling by  $\phi$-mesons could be considered as a signature of  matter dominated by hadronic interactions.

\section{Detectors and Data Sets}
The results presented here are based on data collected in the BES
program at
$\sqrt{s_{NN}}$= 7.7, 11.5 and 39 GeV in Au+Au collisions with the
STAR detector with minimum bias trigger~\cite{trigger} in the year of 2010. The Time
Projection Chamber (TPC)
and Time of Flight (TOF) detectors 
with full $2\pi$ coverage were used for particle identification in the
central rapidity ($\it{y}$) region ($|\it{y}|<$ 1.0). Particles are identified from
information of 
specific energy loss as a function of momentum (using TPC) and 
square of mass as a function of momentum (using TOF).
A  cut on vertex radius (defined as
$V_{R}=\sqrt{V_{x}^{2}+V_{y}^{2}}$, where $V_{x}$ and $V_{y}$ are the
vertex positions along the $\it{x}$ and $\it{y}$ directions) $<$ 2 cm  has been used
to reject events from beam pipe interaction. The total number of
minimum bias events analyzed are about 169
million for 39 GeV, 10.5 million for 11.5 GeV and 4 million for 7.7 GeV.

\section{Flow Analysis Methods}
The Event Plane method~\cite{method} (both full and $\eta$-sub event
plane) has been used for the flow analysis. In this
method each particle correlates with event plane determined from all 
particles in a events except the particle of interest. The event  plane angle $\psi_{2}$ is defined by the equation \\
\begin{equation}
\psi_{2} = \frac{1}{2} \tan^{-1}\frac{\sum  w_{i}\rm{sin}(2\phi_{i})}{\sum
  w_{i}\rm{cos}(2\phi_{i})},
\end{equation}
where sum goes over all the particles  used in  the event plane
calculation. $\phi_{i}$ is the azimuthal angle of the $i^{\rm{th}}$
particle and $w_{i}(=w_{\phi}*w_{pt})$ are weights. The weight $w_{\phi}$, inverse of azimuthal distribution of
particles, has been used to make the distribution of event
planes isotopic in the laboratory system~\cite{method}. In order to improve the event plane resolution, the
weights $w_{pt}$ are set equal to $p_{T}$ up to 2 $\rm{GeV}/\it{c}$ and then
constant at 2.0 above $p_{T} >$ 2 $\rm{GeV}/\it{c}.$ \\
The observed $v_{2}^{\rm{obs}}$ is the second harmonic of the azimuthal
distribution of particles with respect to $\psi_{2}$ \\
\begin{equation}
v_{2}^{\rm{obs}} = <\rm{cos}[2(\phi - \psi_{2})]>,
\end{equation}
where angular brackets denote an average over all particles in all events. Since
finite multiplicity limits the resolution in estimating the angle of
the reaction plane, the observed $v_{2}^{obs}$ has to be corrected for the event
plane resolution as \\
\begin{equation}
v_{2} = \frac{v_{2}^{\rm{obs}}}{<\rm{cos}[2(\psi_{2}-\psi_{r})]>},
\end{equation}
where $\psi_{r}$ is the reaction plane angle. The event plane
resolution~\cite{method} is estimated by the correlation of the events planes of two
sub-events A and B and is given by 
\begin{equation}
<\rm{cos}[2(\psi_{2}-\psi_{r})]> = \it{C} <\rm{cos}[2(\psi_{2}^{A} - \psi_{2}^{B}]>,
\end{equation}
where $\it{C}$ is a constant calculated from the known multiplicity
dependence of the resolution.\\
In $\eta$-sub event plane method~\cite{method} , one defines the event flow vector for each
particle based on particles measured in the opposite hemisphere in
pseudorapidity:
\begin{equation}
v_{2}(\eta_{\frac{+}{}}) =
\frac{<\rm{cos}[2(\phi_{\eta_{\frac{+}{}}} -
  \psi_{2,\eta_{\frac{}{+}}})]>}{\sqrt{<\rm{cos}[2(\psi_{2,\eta_{+}}
    - \psi_{2,\eta_{-}})]>}}.
\end{equation}
Here $\psi_{2,\eta_{+}} ( \psi_{2,\eta_{-}} )$ is the second harmonic
event plane angle defined for particles with positive(negative) pseudorapidity.
An $\eta$ gap of $|\eta| <$ 0.075 between positive and negative
pseudorapidity sub-events has been introduced to suppress non-flow
effects. Event by event resolution correction has been done.
Typical values of event plane resolution ( in $\eta$-sub event plane method ) for minimum bias collision
are 0.43, 0.32 and 0.27 at $\sqrt{s_{NN}}$=39, 11.5 and 7.7 GeV respectively. \\
\section{Results}
$\phi$-mesons are identified using the invariant mass technique from
their decay  to $K^{+} + K^{-}$. Mixed event
technique has been used for combinatorial background subtraction. Figure 1(a) shows
 $\phi$-mesons signal after combinatorial background subtraction in Au+Au collision
  at $\sqrt{s_{NN}}$=39 GeV for $p_{T}$ window 1.1 to 1.3 $\rm{GeV}/\it{c}$ and for
  $0$-$80$$\%$ centrality. The $\phi$-mesons signal fitted with Briet-Wigner
  function and 1st order polynomial for residual background to extract
  $\phi$-mesons yield. The yield distribution as a function of $\phi
  -\psi_{2}$ was fitted by the function
\begin{equation}
A(1+2v_{2}^{\rm{obs}}\rm{cos}[2(\phi-\psi_{2})]),
\end{equation}
to extract the $v_{2}^{\rm{obs}}$ value, where A is a constant. A typical
result for $p_{T}$ window 1.1 to 1.3 $\rm{GeV}/\it{c}$ at $\sqrt{s_{NN}}$=39 GeV is
shown in figure 1(b).
\begin{figure}[h]

%\centerline{\includegraphics[scale=0.4]{39GeV_ncq_particle.eps}}
\includegraphics[scale=0.285]{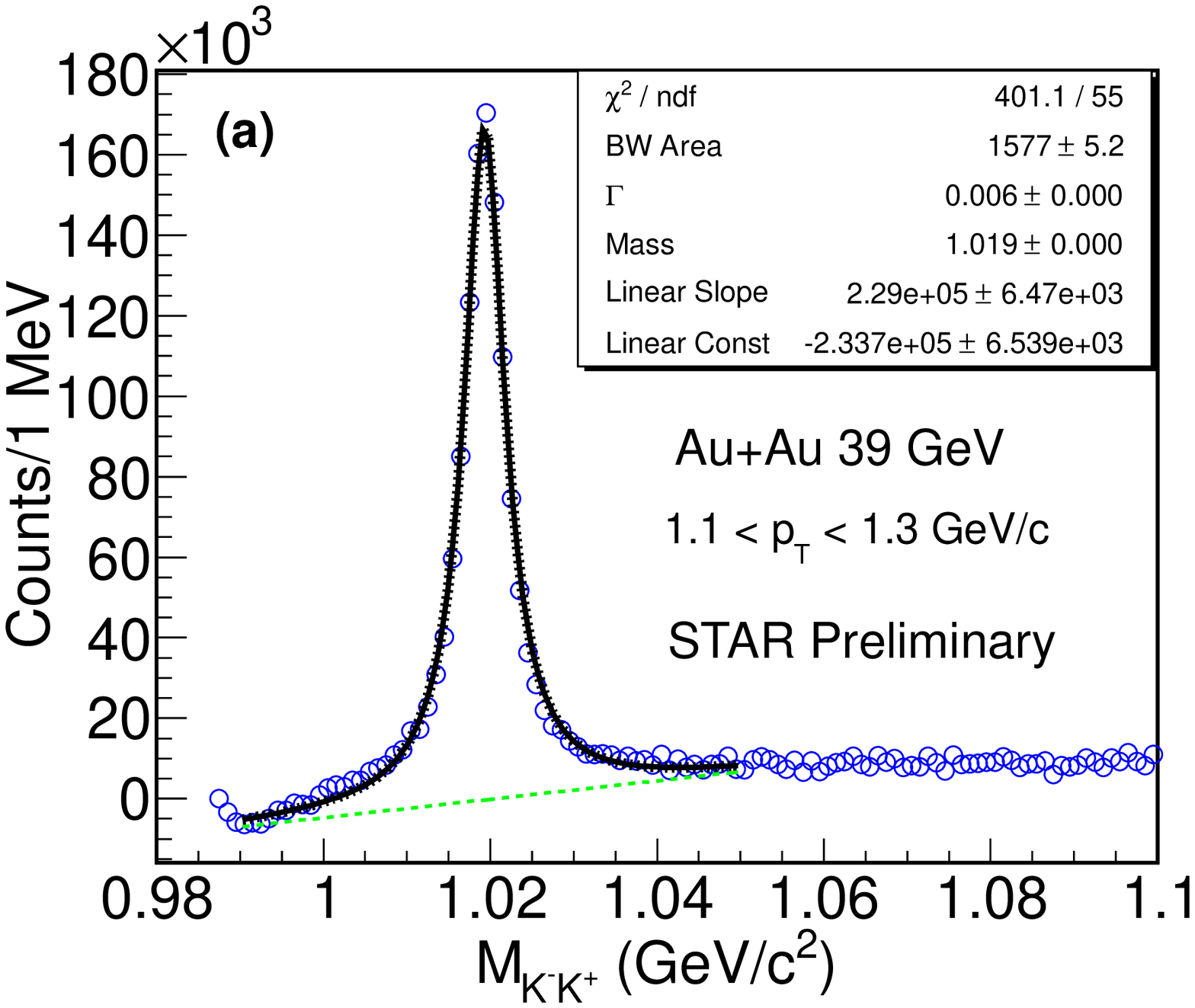}
\includegraphics[scale=0.28]{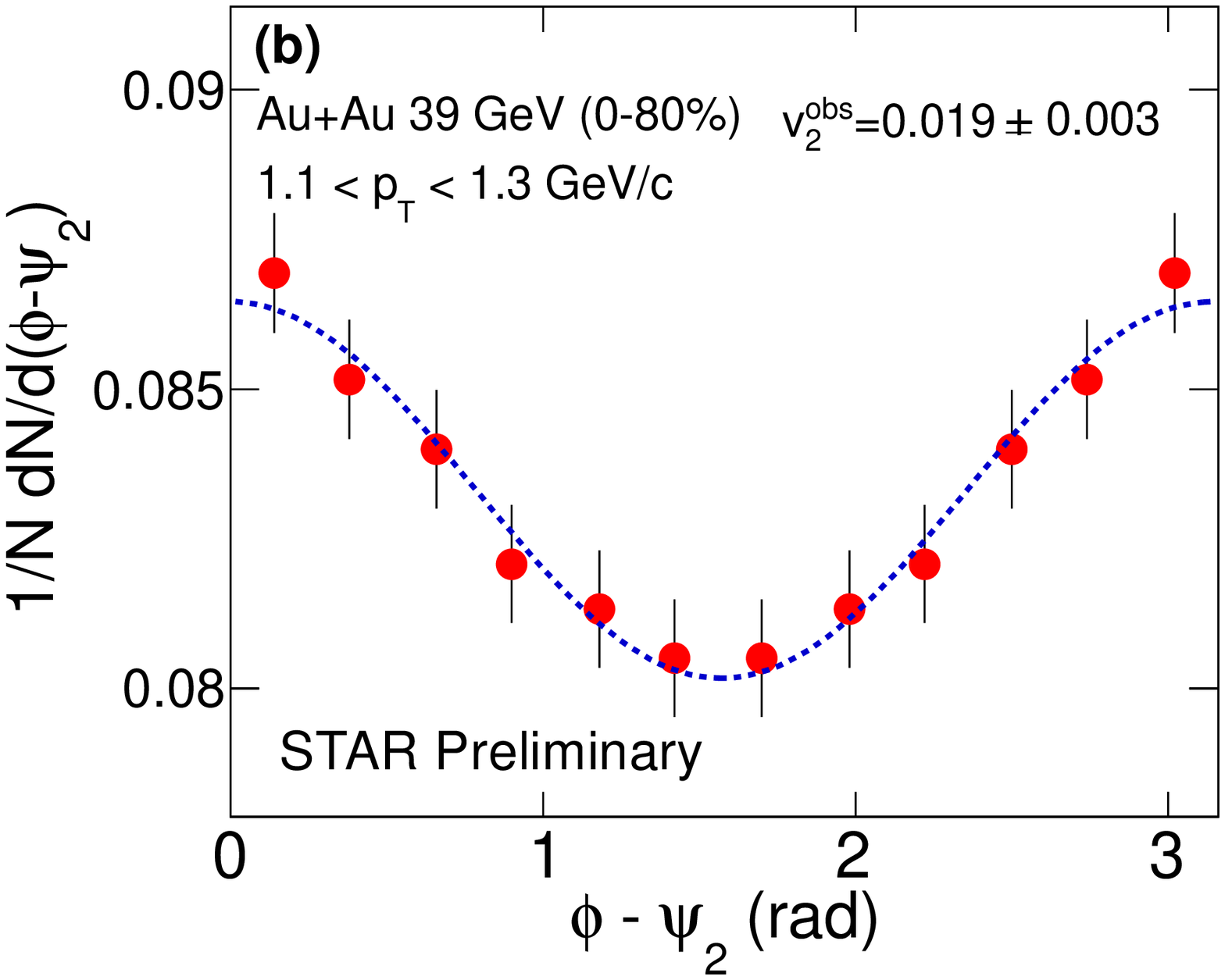}

\caption{ (a) $\phi$-mesons signal after combinatorial background subtraction in Au+Au
  collision (0-80$\%$) at $\sqrt{s_{NN}}$= 39 GeV from a selected
  $p_{T}$ bin (1.1 to 1.3 $\rm{GeV}/\it{c}$). (b) $\phi-\psi_{2}$ distribution
for $\phi$-meson at $1.1 < p_{T} <1.3$ $GeV/\it{c}$ in Au+Au collision at
$\sqrt{s_{NN}}$=39 GeV }
\end{figure} 
  
The  measurements  of $\phi$-meson $v_{2}$ as function of
transverse momentum at $\sqrt{s_{NN}}$= 39, 11.5 and 7.7 GeV  is
shown in figure 2. These results are from $\eta$-sub event plane
method.  The published results~\cite{ncq2} at
$\sqrt{s_{NN}}$=200 GeV in Au+Au collision 
\begin{figure}[h]
\centerline{\includegraphics[scale=0.35]{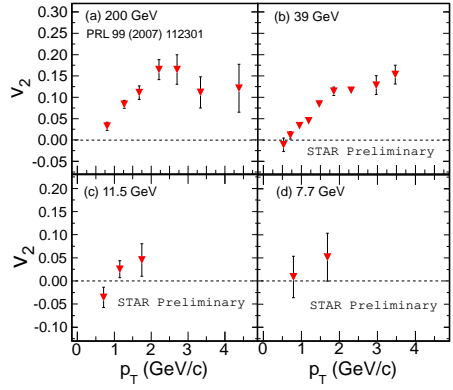}}
 \caption{$v_{2}$ vs $p_{T}$ of $\phi$-meson for Au + Au collision (0-80$\%$) at
   $\sqrt{s_{NN}}$= 7.7, 11.5, 39 and 200 GeV with $|\it{y}| < $ 1.0 . }
\end{figure} 
  is also shown in figure 2 (a) for comparison . At $\sqrt{s_{NN}}$=
11.5 GeV , $\phi$-meson $v_{2}$ is observed to be significantly smaller than
39 and 200 GeV.  Due to limited statistics there are only two data
points with large statistical error for $\phi$-meson $v_{2}$ at
$\sqrt{s_{NN}} $= 7.7 GeV.  To understand these results we will
discuss effect partonic and hadronic  interaction  on $\phi$-mesons
$v_{2}$. The two main possibility of $\phi$-meson production are (a) kaon
coalescence and (b) coalescence of s and $\bar{s}$ quarks in the medium. The recent
results~\cite{result1} from RHIC and NA49 Collaboration~\cite{result2}
to $\phi$-meson production 
shows that the contribution from  kaon
coalescence should be small in this energy range and  the
$\phi$-mesons production is dominated by partonic interaction. So the
contribution to $\phi$-meson $v_{2}$ is mostly
from partonic phase. 

\begin{figure}[!h]
%\centerline{\includegraphics[scale=0.4]{39GeV_ncq_particle.eps}}
\includegraphics[scale=0.28]{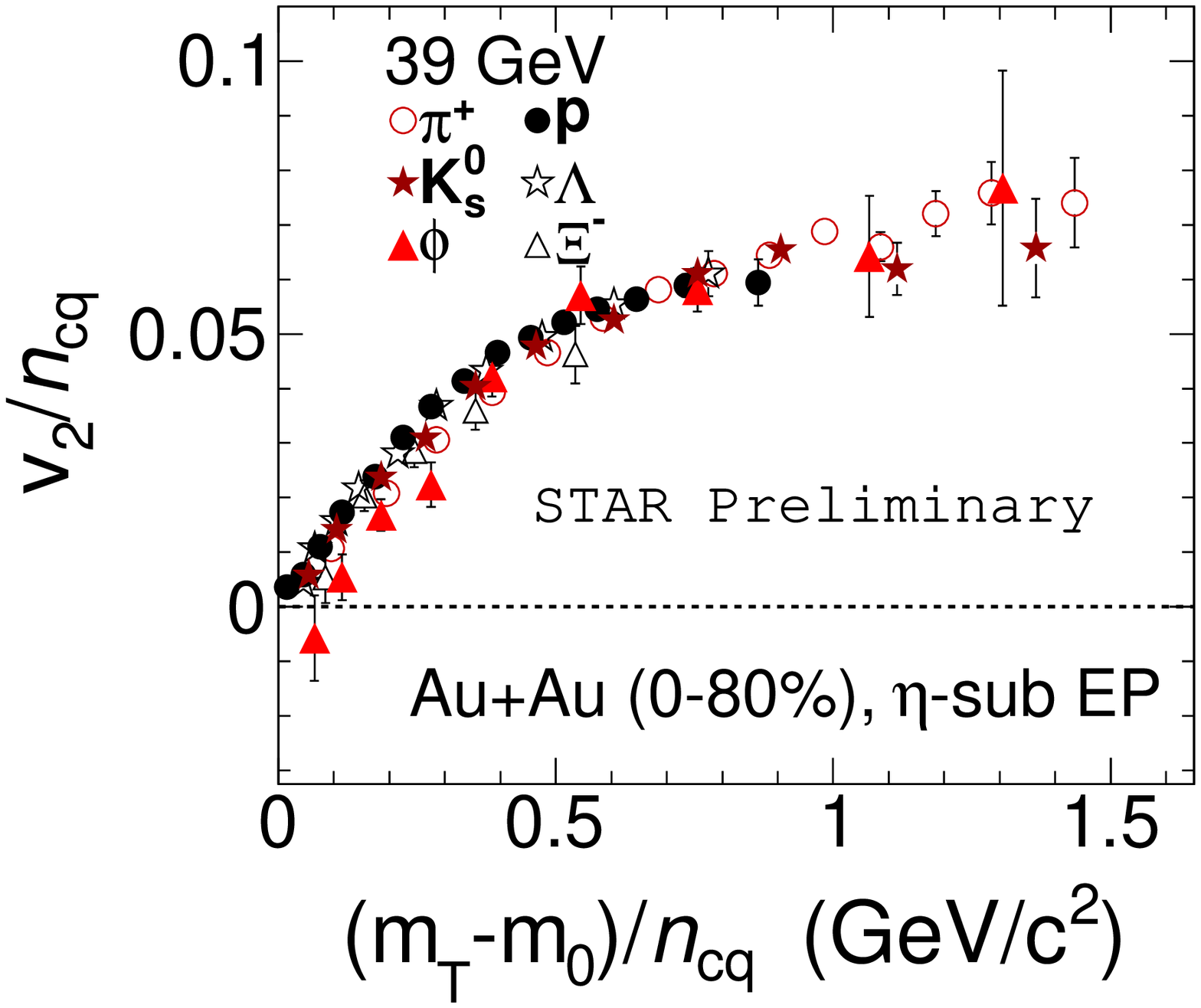}
\includegraphics[scale=0.28]{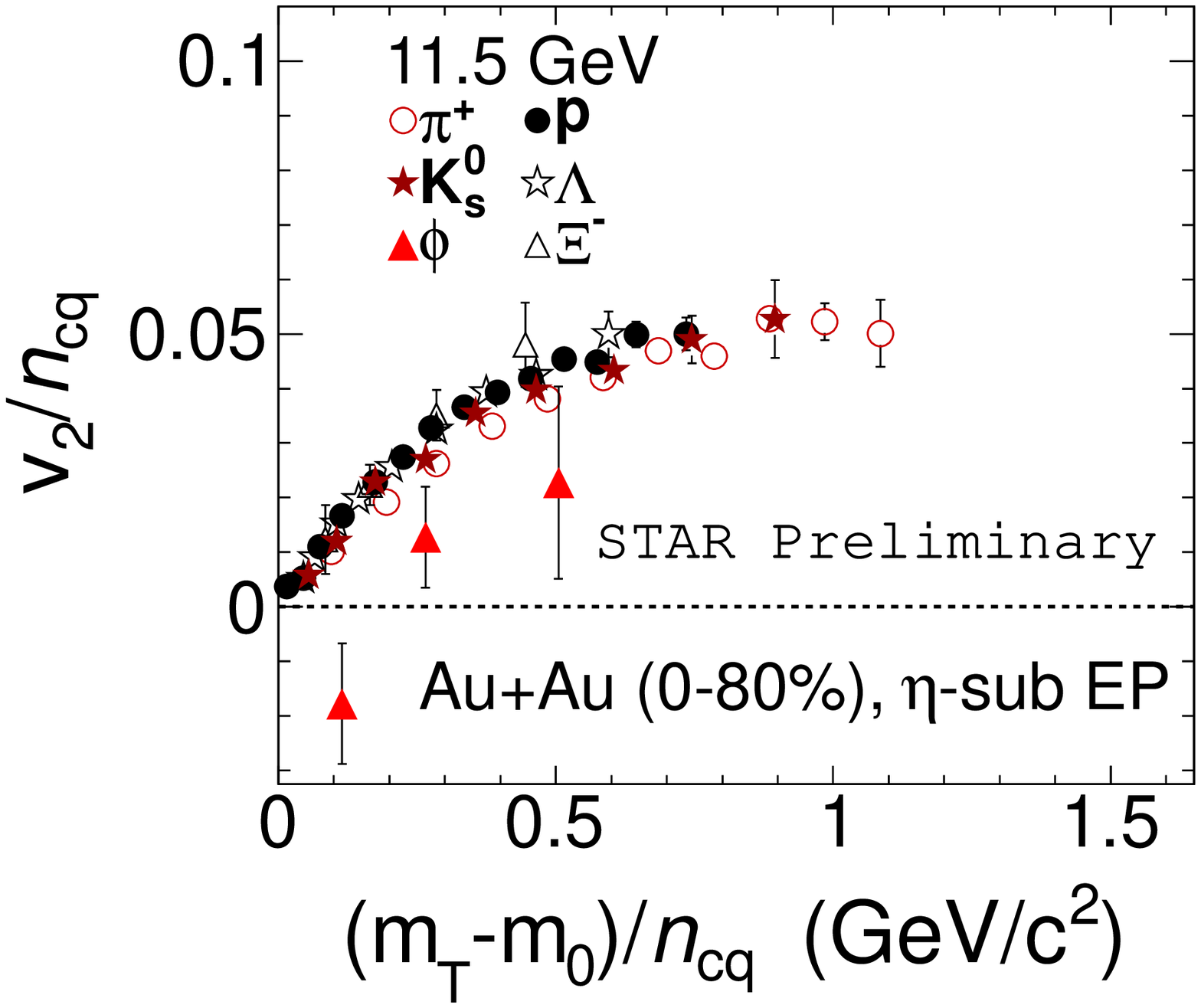}

\caption{ $v_{2}/n_{\rm{cq}}$ vs $(m_{T} -m_{0})/n_{\rm{cq}}$ in Au + Au
collisions at $\sqrt{s_{NN}}$= 11.5 and 39 GeV.}
\end{figure} 

Now we will discuss the effect of system
evolution on $v_{2}$ of $\phi$ mesons. Phenomenological analysis~\cite{smallx}
and experimental result on coherent $\phi$ photo-production~\cite{result4} suggest
that interaction cross-section of $\phi$-mesons with other hadrons
is much smaller than that of other particles. Due small hadronic interaction
cross-section of $\phi$-mesons, they freeze out very early and close to
chemical freeze out temperature. So the effect of hadronic interactions
on $\phi$-mesons $v_{2}$ is very small and most of the contribution on
$v_{2}$ is from partonic phase. Therefore large  $\phi$-meson $v_{2}$
indicates the formation partonic matter and small $v_{2}$ of could
indicate dominance of hadron interactions.      \\
Figure 3 shows $v_{2}$ divided by number of constituent quark  as
function  of  $(m_{T} - m_{0})/n_{\rm{cq}}$, where $m_{T} ( =\sqrt{p_{T}^{2}
+ m_{0}^{2}})$ is the transverse mass, $p_{T}$ is the transverse
momentum, and $m_{0}$ is the mass of the hadron, at $\sqrt{s_{NN}}$ =
39 and 11.5 GeV. The $\phi$-mesons $v_{2}$ shows similar $v_{2}$
values as
other hadrons at $\sqrt{s_{NN}}$= 39 GeV  whereas for $\sqrt{s_{NN}}$=
11.5 GeV $\phi$-mesons  falls off the trend from the other
hadrons. The mean deviation of $\phi$-mesons $v_{2}$ from the
$\pi$-meson $v_{2}$ is 2.6 $\sigma$ ($\sigma$ is the error on
$\phi$-meson $v_{2}$).The NCQ scaling has been
understood by considering quark coalescence as mechanism of
hadronization and it is believed to be indication of
deconfinement~\cite{result5}. The study of $\phi$-mesons $v_{2}$ using A Multi Phase
Transport Model (AMPT) shows that,  partonic interaction are necessary
for NCQ scaling of $v_{2}$ of hadrons~\cite{result6}. So the small 
magnitude of $\phi$-meson $v_{2}$ and its deviation from the values of
other hadrons  at
$\sqrt{s_{NN}}$=11.5 GeV could be effect of a matter where hadronic
interactions are dominant.   

\section{Summary}
We have presented $v_{2}$ of $\phi$-mesons in Au + Au
collision at $\sqrt{s_{NN}}$= 7.7, 11.5 and 39 GeV  obtained by the
STAR experiment. Large $\phi$-mesons $v_{2}$ at $\sqrt{s_{NN}}$=39 GeV
indicates that partonic collectivity has been developed at $\sqrt{s_{NN}}$ =
39 GeV as it has been seen
before at top RHIC energies. Different trend has been observed for
$\phi$-mesons $v_{2}$ at $\sqrt{s_{NN}}$=11.5 GeV. The deviation of
$\phi$-meson $v_{2}$ from other hadrons could 
 indicate the dominance of hadronic interactions with decreasing the
 beam energy.  

 \section{ACKNOWLEDGMENTS}
Financial support from Board of Research on Nuclear Science
 (project sanction No. 2010/21/15-BRNS),
Government of India is gratefully acknowledged.


\begin{thebibliography}{99}

\bibitem{hydro} P.F. Kolb et al. Nucl. Phys. A715, (2003) 653c

\bibitem{smallx} J. Rafelski and B. Muller, Phys. Rev. Lett. 48 (1982)
  1066;
\bibitem{ncq1} D. Molnar and S. A. Voloshin, Phys. Rev. Lett. 91
  (2003) 092301
\bibitem{ncq1a} J. Adams et al. ( STAR Collaboration) Phys. Rev. Lett. 92 (2004) 052302 
\bibitem{ncq2} B. I. Abelev et al. (STAR Collaboration) Phys. Rev. Lett. 99 (2007) 112301

\bibitem{trigger} B. I. Abelev et al. ( STAR Collaboration)
  Phys. Rev. C 81 (2010) 024911  
\bibitem{method} A. M. Poskanzer and S. A. Voloshin,  Phys. Rev. C 58
  (1998) 1671

\bibitem{result1} B. I. Abelev et al. (STAR Collaboration)
  Phys. Rev. C 79 (2009) 064903

\bibitem{result2} NA49 Collaboration, C. Alt et al.,Phys. Rev. C 78
  (2008) 044907
%\bibitem{result3}
\bibitem{result4} A. Sibirtsev et al. Eur. Phys. J. A 29 (2006) 209.
\bibitem{result5} S. Pratt and Subrata Pal, Phys. Rev. C 71, (2005) 014905
\bibitem{result6}  B. Mohanty and N. Xu , J. Phys. G 36, (2009) 064022


\end{thebibliography}
\end{document}